\documentclass[aps,prl,twocolumn,superscriptaddress]{revtex4}
\usepackage{amsmath,epsfig,mathrsfs,graphicx,amssymb}

\begin{document} 
\title{Noninvasive Quantum Measurement of Arbitrary Operator Order by Engineered Non-Markovian Detectors}
\author{Johannes B{\"u}lte}
\affiliation{Fachbereich Physik, Universit{\"a}t Konstanz, D-78457 Konstanz, Germany}
\author{Adam Bednorz}
\affiliation{Faculty of Physics, University of Warsaw, ul. Pasteura 5, PL02-093 Warsaw, Poland}
\author{Christoph Bruder}
\affiliation{Department of Physics, University of Basel, Klingelbergstrasse 82, CH-4056 Basel, Switzerland}
\author{Wolfgang Belzig}
\affiliation{Fachbereich Physik, Universit{\"a}t Konstanz, D-78457 Konstanz, Germany}

\date{\today}

\begin{abstract}
  The development of solid-state quantum technologies requires the
  understanding of quantum measurements in interacting, non-isolated
  quantum systems. In general, a permanent coupling of detectors to a
  quantum system leads to memory effects that have to be taken into
  account in interpreting the measurement results. We analyze a
  generic setup of two detectors coupled to a quantum system and
  derive a compact formula in the weak-measurement limit that
  interpolates between an instantaneous (text-book type) and almost
  continuous - detector dynamics-dependent - measurement. A quantum
  memory effect that we term \textit{system-mediated detector-detector
  interaction} is crucial to observe non-commuting observables
  simultaneously. Finally, we propose a mesoscopic double-dot detector
  setup in which the memory effect is tunable and that can be used to
  explore the transition to non-Markovian quantum measurements
  experimentally.
\end{abstract}

\maketitle

Historically the interpretation of quantum measurement uses the projection
postulate implying an instantaneous collapse of the wave function in
the measurement process \cite{neu32}. Whereas this scheme fits
e.g. ideal photo-detection very well, it is unsuitable in most other
measurement procedures. For instance, in a measurement of a current in
a solid-state environment, the system-detector interaction is much
weaker and a collapse is avoided. A great deal of theoretical and
experimental investigations have been carried out since von Neumann's
work \cite{neu32,Braginsky:92,wm09}. In particular, the limit of
noninvasive measurement processes has been studied theoretically
\cite{aharonov1988} as well as experimentally \cite{katz:08,vijay:12}

In mesoscopic physics, the question of current noise in the quantum
regime \cite{blanter:00} has attracted a lot of of interest over the
years. In particular, the question of measuring the current correlator
was addressed early on
\cite{lesovik:96,LeeLevitov96,aguado:00,gavish:04,PhysRevB.81.125112}. While
most experiments address high-frequency correlations in agreement with
the symmetrized correlator
\cite{zakka:07,gabelli:08,palacios-laloy:10}, with on-chip detectors
the quantum non-symmetrized noise can be extracted
\cite{deblock:03,gustavsson:07,basset:12,ferraro:14}.  First
theoretical strides towards the measurement and interpretation of
general unsymmetrized operator orders have been reported
\cite{beenakker:01,salo:06,lebedev:10,plimak:12,bednorz:10}.  A
phenomenological approach showing that memory effects allow to access
nonsymmetrized correlation functions has been investigated e.g. in
\cite{BBRB13}, but a specific and realistic treatment was lacking.

The goal of this Letter is to understand which correlations are
obtained in a concrete quantum measurement setup, in which two
detectors coupled weakly and continuously to a system are read out
independently.  We find that the measurement outcomes can be expressed
by noise and response functions of the system and the
detectors. Hence, these outcomes depend crucially on the internal
dynamics of the detectors and, therefore, by suitably engineered
detectors the measurement can be tuned such that a specific operator
order is measured.  We propose a mesoscopic double-dot detector setup
to explore the transition to non-Markovian quantum measurements
experimentally. Going beyond the usually discussed bath-induced
non-Markovian self-interaction \cite{breuer:02}, we identify the
\textit{system-mediated detector-detector interaction} as crucial
ingredient to observe non-commuting observables simultaneously.

\begin{figure}[t]
    \centering
    \includegraphics[width=7cm]{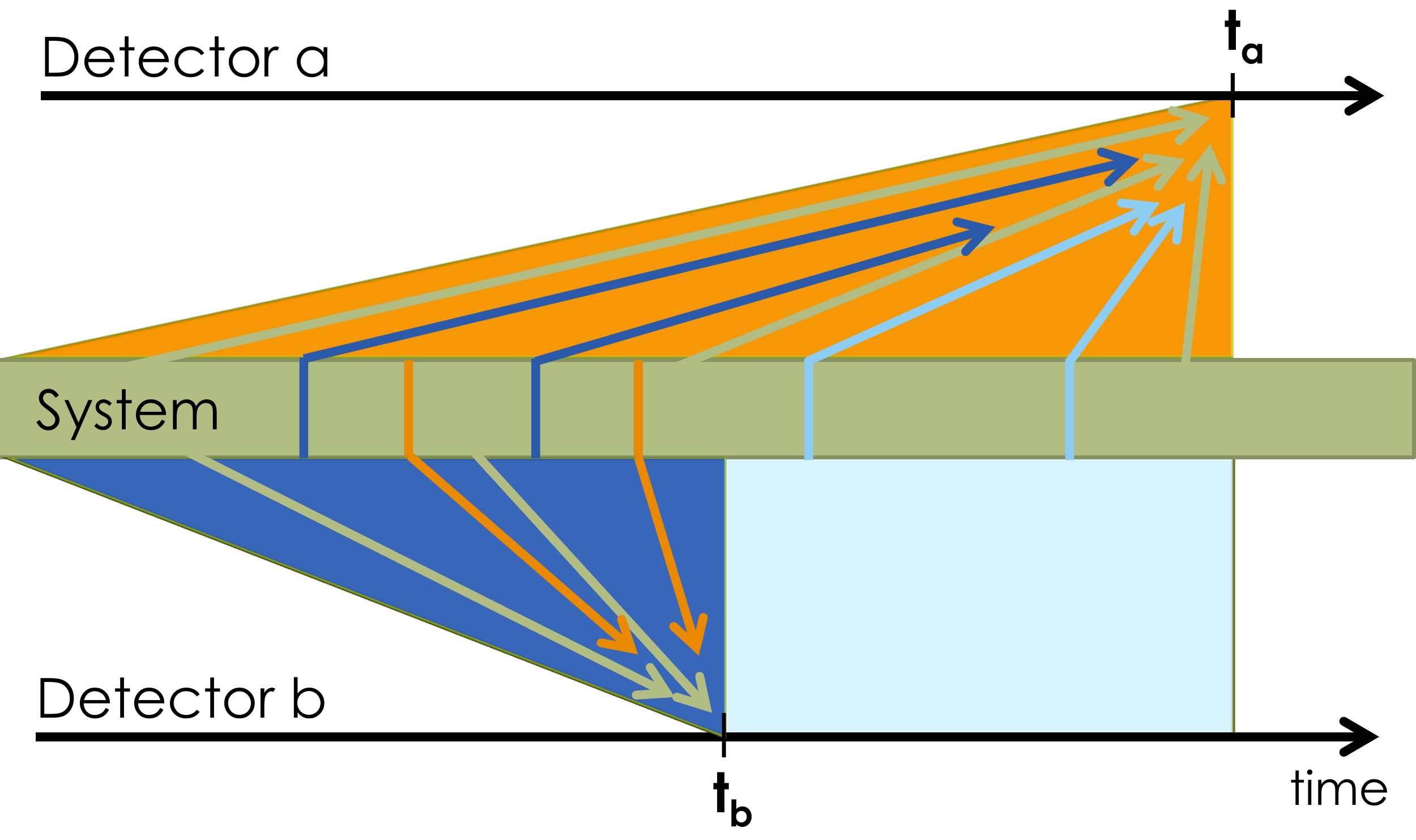}
    \caption{Non-Markovian cross-correlation measurement: Two detectors $a$ and $b$ are coupled to a system for a finite time and read out at times $t_a > t_b$. Detector $b$ collects information directly on the system and on detector $a$ (which are transmitted through the system).
The presence of detector $b$ after its readout (light blue) also contributes to the information which detector $a$ collects on detector $b$ at a later time $t_a$.}
     \label{fig:detectionscheme}
\end{figure}

In the following, we will assume that two detectors are coupled to a
common system by small interaction Hamiltonians, which allows us to
linearize the time-evolution and obtain a microscopic expression for
the observed quantum correlation. The measurement procedure is
illustrated in Fig.~\ref{fig:detectionscheme}: Besides the direct
influence of the system onto the detector observables (green arrows),
the detectors influence \textit{each other} mediated by the system
(blue and orange arrows). This influence depends on the time-dependent
response of the system and will lead to interesting consequences.  We
start by deriving the second-order cross correlation for a
non-Markovian weak quantum measurement from a microscopic model. We
will show that the different contributions can be understood in terms
of a linear-response interaction of the subsystems involved. After
making some general statements on the properties and relevance of
terms beyond the textbook approach of symmetric operator order, we
will illustrate our findings in several examples.

{\em Microscopic model.} --- The Hamiltonian of the system
to be measured is denoted as
$\hat H_0$, and we will consider two independent detectors described
by $\hat H_a$ and $\hat H_b$ that are coupled to the system by $\hat H_{\text{int}}$. The total Hamiltonian is given by
\begin{equation}
\hat  H=\sum_{\alpha=0,a,b} \hat H_\alpha+\hat H_{\text{int}}\:.
\end{equation} 
The two detectors with detector variables $\hat D_a$ and $\hat D_b$ are weakly and linearly coupled to the system observables $\hat A$ and $\hat B$,
\begin{equation}
 \hat H_{\text{int}}=\eta_a\hat D_{a}\hat A + \eta_b\hat D_{b}\hat B\:,
\label{Hint}
\end{equation} 
where $\eta_a$ and $\eta_b$ denote small coupling parameters which in general can be modulated in time.
There are no restrictions on the detector Hamiltonians $\hat H_{\alpha}$.

\begin{figure}[t]
    \centering
    \includegraphics[width=7cm]{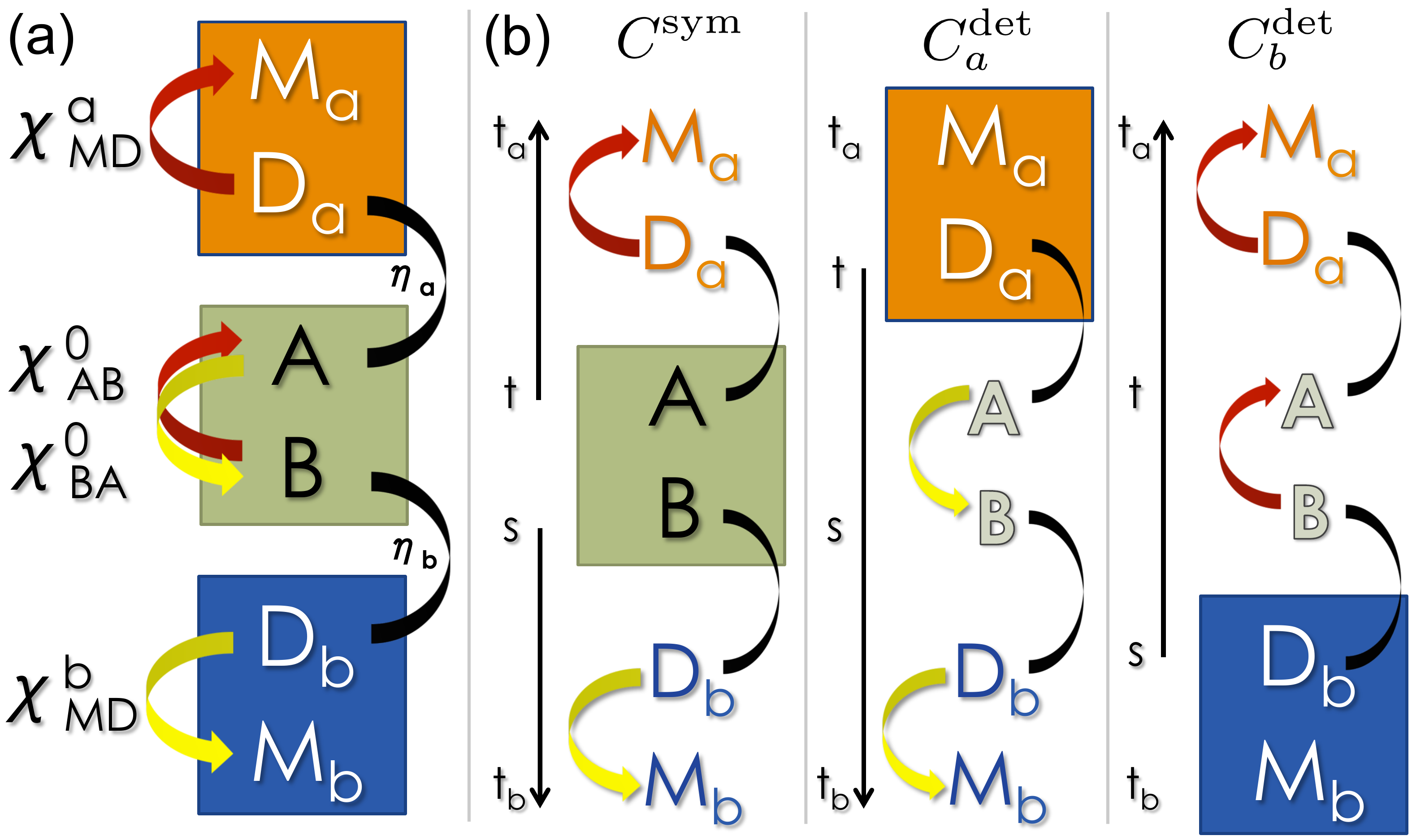}
    \caption{(a) Couplings and correlations between the different variables. The inter-system couplings $\eta_{\alpha}$ connect the detector variables $\hat D_\alpha$ to the system variables $\hat A$ and $\hat B$ at equal times. 
Within the system and the detectors information is transmitted  non-locally in time via the response functions $\chi$.
(b) Schematic representation of the second-order processes which
contribute to the cross-correlation measurement. The straight arrows indicate the direction and causality of the time flow.}
     \label{fig:secondorderprocesses}
\end{figure}

In the following we will study the correlation measurement scheme
shown in Fig.~\ref{fig:detectionscheme}. The density matrix of the
total system is assumed to be in a product state $\hat
\rho=\hat\rho_0\hat\rho_a\hat\rho_b$ at the initial time $t_0$ (not shown). The
measurement/readout procedure of the detectors is described by Kraus
operators $\hat K_{m_i}$~\cite{wm09}.
The first measurement ends with the readout of detector $b$ at time $t_b$ with result $m_b$ and the density matrix
$\hat \rho'=\frac{\hat K_{m_b}\hat U(t_b)\hat\rho_0\hat\rho_a\hat\rho_b\hat U^\dagger(t_b)\hat K_{m_b}^\dagger}{\text{Tr}\left\{\hat K_{m_b}\hat U(t_b)\hat\rho_0\hat\rho_a\hat\rho_b\hat U^\dagger(t_b)\hat K_{m_b}^\dagger\right\}}$.
The unitary time evolution operator can be separated as $\hat U(t)=\hat
U_0(t)\hat U_\text{I}(t)$. Here, $\hat U_0(t)=e^{-\frac{i}{\hbar}(\hat H_0+\hat
  H_{a}+\hat H_{b})(t-t_0)}$ generates the time evolution in the
uncoupled subsystems and $\hat U_\text{I}(t)~=~\mathcal{T}e^{-\frac{i}{\hbar}\int^t_{t_0} \hat
  H_{\text{int}}^\text{I}(t') \text{d}t'}$ describes the time evolution in
the interaction picture induced by $\hat H_{\text{int}}$. Here,  $\mathcal{T}$ denotes the time-ordering operator. The second
measurement ends with the readout of detector $a$ at time $t_a\geq
t_b$ with result $m_a$. The (unnormalized) total state after this procedure is given by
$\hat \rho''=\hat K_{m_a} \hat U(t_a-t_b)\hat\rho'\hat U^\dagger(t_a-t_b)\hat K_{m_a}^\dagger$. Its trace Tr$\left\{\hat \rho''\right\}=p(m_a,t_a|m_b,t_b)$ gives the conditional probability of finding $m_a$ if $m_b$ has been measured before; the unconditional probability is obtained by applying Bayes' theorem $p(m_a,t_a;m_b,t_b)=p(m_a,t_a|m_b,t_b)p(m_b,t_b)$.
To establish the relation between measurement results of the
two detectors and operator order of the
corresponding observables we consider
\begin{align}
&C(t_a,t_b) = \int\text{d}m_a\int\text{d}m_b\, m_a\, m_b\, p(m_a,t_a;m_b,t_b)\nonumber\\
&-\int\text{d}m_a \, m_a\, p(m_a,t_a)\int\text{d}m_b\, m_b\, p(m_b,t_b).
\label{exactcorrelator}
\end{align}
Using the weak-coupling assumption, expanding $\hat U_I$ to second
order in $\eta$ and rescaling $C\to C/\eta_a\eta_b$, we find that the correlation function can be written as the sum of three contributions
\begin{equation}
 C(t_a,t_b)= C^{\text{sym}}(t_a,t_b)+C^{\text{det}}_{a}(t_a,t_b)+C^{\text{det}}_{b}(t_a,t_b)\:,
\label{correlator_time}
\end{equation}
with
\begin{align}
C^{\text{sym}}(t_a,t_b)=&\int\text{d}t\,\text{d}s\,\chi^a_{MD}(t_a,t)\chi^b_{MD}(t_b,s)S_{AB}^{0}(t,s),\nonumber\\
C^{\text{det}}_{a}(t_a,t_b)=&\int\text{d}t\,\text{d}s\,\chi^b_{MD}(t_b,s)\chi^0_{BA}(s,t)S_{MD}^{a}(t_a,t),\nonumber\\
C^{\text{det}}_{b}(t_a,t_b)=&\int\text{d}t\,\text{d}s\,\chi^a_{MD}(t_a,t)\chi^0_{AB}(t,s)S_{MD}^{b}(t_b,s).
\label{Cts}
\end{align}
Here we introduced the response functions
\begin{equation}
 \chi^\alpha_{XY}(t,t')=-\frac{i}{\hbar}\theta(t-t')\langle[\hat X_\alpha(t),\hat Y_\alpha(t')]\rangle_\alpha\:,
\label{response}
\end{equation} 
and the symmetrized noise functions
\begin{equation}
 S^\alpha_{XY}(t,t')=\frac{1}{2}\langle\{\delta \hat X_\alpha(t),\delta \hat Y_\alpha(t')\}\rangle_\alpha\:.
\label{noise}
\end{equation} 
The expectation values in Eqs.~(\ref{response}) and (\ref{noise}) are
calculated with the initial system density matrix $\hat\rho_0$
respectively the density matrices
$\hat\rho_{\alpha}(t)=\theta(t_{\alpha}-t)\hat\rho_{\alpha}+\theta(t-t_{\alpha})\hat\rho'_{\alpha}$
of the detectors ($\alpha=a,b$),
which can incorporate a change of the detector state from
$\hat\rho_{\alpha}$ to $\hat\rho_{\alpha}'$ due to its readout at time
$t_{\alpha}$. The detector observable is given by
$\hat M_{\alpha}=\int\text{d}m_\alpha\, m_\alpha \hat
K_{m_\alpha}^{\dagger}\hat K_{m_\alpha}$.

Equations~(\ref{correlator_time}) and (\ref{Cts}) are the main result
of our Letter. The origin of the various terms is illustrated in
Fig.~\ref{fig:secondorderprocesses}. $C^{\text{sym}}$ describes the
symmetrized correlation of the system observables $\hat A$ and
$\hat B$ which is transmitted to both detectors via $\chi^a$ and
$\chi^b$. $C^{\text{det}}_{a}$ corresponds to the internal correlation
of $\hat M_a$ and $\hat D_a$ in detector $a$ where $\hat D_a$ is
measured by detector $b$ via $\chi^0$ and $\chi^b$ (and similarly for
$C^{\text{det}}_{b}$).
The system takes a new role as a linear-response mediator here, since the commutators in the response function
Eq.~(\ref{response}) make information accessible that is not contained
in the symmetrized noise, Eq.~(\ref{noise}).   

If system and detectors are initialized in stationary states,
Eqs.~(\ref{response}) and (\ref{noise}) depend only on the time differences
and Eq.~(\ref{correlator_time}) can be expressed in Fourier space as $C(\omega)=\int{\text{d}(t_a-t_b)}e^{i\omega(t_a-t_b)}C(t_a,t_b)$ with the result
\begin{align}
C^{\text{sym}}(\omega)=&\chi^a_{MD}(\omega)\chi^b_{MD}(-\omega)S^0_{AB}(\omega),\nonumber\\
C^{\text{det}}_{a}(\omega)=&\chi^b_{MD}(-\omega)\chi^0_{BA}(-\omega)S^{a}_{MD}(\omega),\nonumber\\
C^{\text{det}}_{b}(\omega)=&\chi^a_{MD}(\omega)\chi^0_{AB}(\omega)S^{b}_{MD}(-\omega)\:.
\label{Somega}
\end{align}
For simplicity we concentrate on the second-order cross-correlation here, the generalization is discussed in \cite{SM}.
Hence, in the frequency domain
and for equal detectors we have $C^{\text{sym}}(\omega)=|\chi_{MD}(\omega)|^2
S_{AB}^{0}(\omega)$. Thus, the non-Markovian nature of our setup
results in a simple frequency filter effect for the symmetric part of the measurement.

We will now focus on the system-mediated detector-detector
interactions $C^{\text{det}}_{\alpha}$ and discuss their relevance and properties, especially
their effect on the finally measured operator order. Since the
detectors measure each other's noise in linear response through the
system, the presence of the commutator in the response functions leads
to the appearance of antisymmetrically ordered system terms in
addition to the symmetrized expressions which can combine in various
ways. We find that
\textit{(i)} $C^{\text{det}}$ is a non-Markovian quantity as it depends on the memory of the detector and the time history of the measurement process.
\textit{(ii)} the response--to--noise ratio $\chi/S$ of a detector is a indicator of the relevance of $C^{\text{det}}$ and therefore the measured operator order.
\textit{(iii)} $C^{\text{det}}_a$ and $C^{\text{det}}_b$ can cancel with clever detector engineering.
\textit{(iv)} For certain systems, $C^{\text{det}}$ leads to observable effects even in the Markovian coupling limit, i.e. for instantaneous interaction of system and detectors.
The observations \textit{(i)-(iv)} are illustrated in the corresponding examples in the following.

\begin{figure}[t]
    \centering
    \includegraphics[width=7cm]{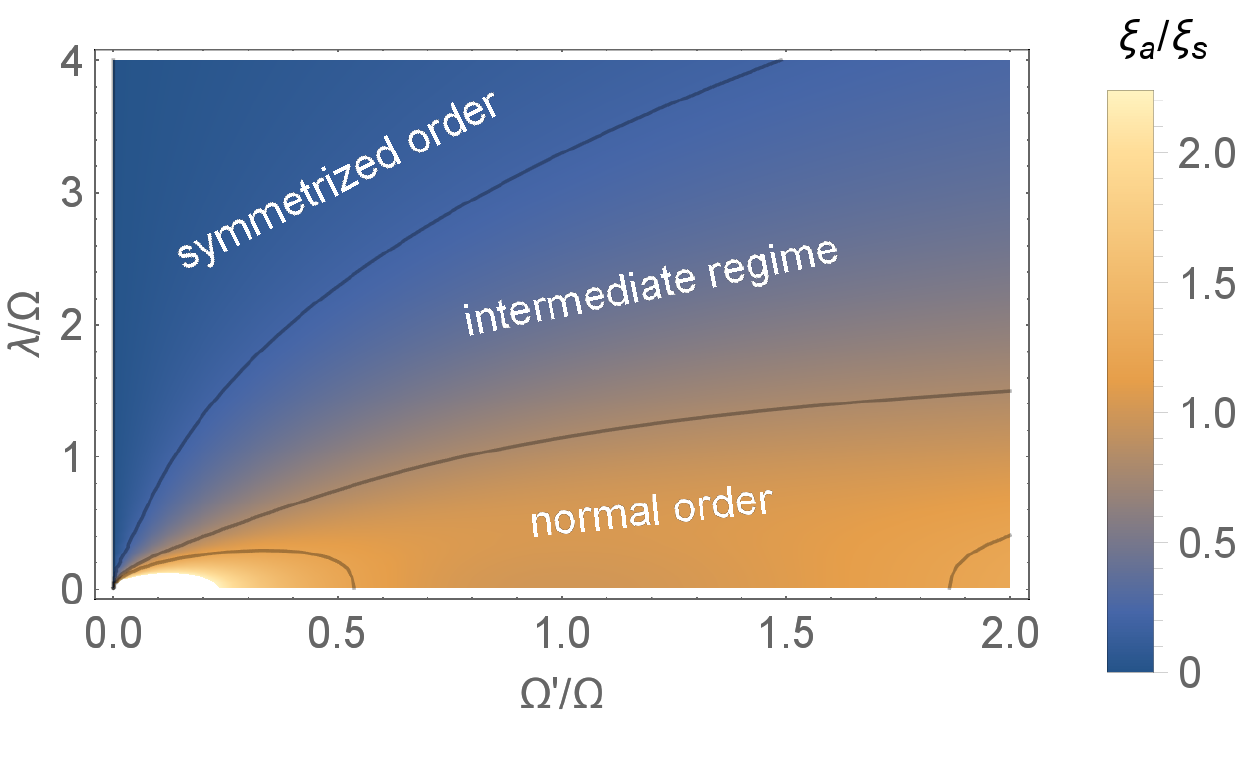}
    \vspace{-0.5cm}
    \caption{Ratio of the weights $\xi_{a}$ and $\xi_{s}$ of the antisymmetric and symmetric contributions to Eq.~(\ref{exosc}) as a function of the detectors' frequency $\Omega'$ and damping $\lambda$ in units of the system frequency $\Omega$ at zero detector temperature.}
     \label{dampedHOxis1}
\end{figure}

\textit{(i) Ladder-operator measurement:}
We study a system that is assumed to be a harmonic oscillator with frequency
$\Omega$. The detectors are assumed to be identical harmonic
oscillators with frequency $\Omega'$, and their memory is controlled
by the damping parameter $\lambda$. This damping results from the
dissipation to a bath, see \cite{SM}. The measurement is taken at
coinciding times $t_a=t_b=0$. We are interested in the correlation of
the ladder operators $\hat a$ and $\hat a^{\dagger}$ of the
system. Since they are not hermitian, we have to perform two
measurements in which the detectors are coupled to the position in one
measurement with the resulting correlation function $C_{xx}$ and to
the momentum in the other measurement resulting in $C_{pp}$. The sum
of these correlators $C=(C_{xx}+C_{pp})/2$ corresponds to the scalar
relation $\alpha\alpha^{*}=(x^2+p^2)/2$ with $\alpha=(x+ip)/\sqrt{2}$
and is given by
\begin{align}
 C=\xi_s\langle\{\hat a,\hat a^\dagger\}\rangle-\xi_{a}\langle[\hat a,\hat a^\dagger]\rangle\:,
\label{exosc}
\end{align}
where ${\xi_{s}=\frac{1}{2}\int \text{d}t\text{d}s \,
  \chi_{MD}^{a}(t)\chi_{MD}^{b}(s)\cos(\Omega (t-s))}$ and
$\xi_{a}=\int\text{d}t\text{d}s\,\theta(s-t)\left[\chi_{MD}^b(t)S_{MD}^a(s)+\chi_{MD}^a(t)S_{MD}^b(s)\right]$
$\times\sin(\Omega (s-t))$ are the weights of the symmetric and
antisymmetric contributions. The antisymmetric contribution is essentially determined by the system-mediated detector-detector interaction $C_{\text{det}}$. Choosing the detector
variable $\hat D=\hat p$ and the detector observable $\hat M~=~\hat x$
yields $\chi_{MD}(t)=\theta(t)e^{-\lambda t}\cos(\Omega' t)$ and
$S_{MD}(t)~=~e^{-\lambda t}\sin(\Omega' t)\coth(\beta\Omega'/2)/2$
with the detectors inverse thermal energy $\beta$.
Figure~\ref{dampedHOxis1} shows the ratio of the weights $\xi_{a}$ and $\xi_{s}$ as a function of $\Omega'$ and $\lambda$.
If the detector oscillators are only slightly damped ($\lambda\ll\Omega'$), i.e. have a long memory, the symmetric and antisymmetric contributions are of the same size and lead to normal order of the operators in the expression for $C\propto\langle\hat a^\dagger\hat a\rangle$. If the detector oscillators are overdamped ($\lambda\gg\Omega'$), i.e. approach the Markovian limit, the antisymmetric contribution vanishes and the correlation function corresponds to symmetrized operator order $C\propto\langle\{\hat a^\dagger,\hat a\}\rangle$.

\textit{(ii) Relevance of the system-mediated contributions:}
The response to noise ratio $\chi/S$ of the detector characterizes the strength
of the asymmetric system operator order, which can be illustrated e.g. with a monochromatic laser beam. $S$ and $\chi$ resemble a harmonic oscillator in many respects, with the difference that the noise is proportional to the photon number $N$ whereas the response is unaffected by it. Thus we have $\chi/S\approx 1/N$. A single photon behaves very much like a harmonic oscillator and $C^{\text{sym}}$ and the $C^{\text{det}}_{\alpha}$ contribute in the same order with a fine tuning possible by damping and resonance effects as shown in Fig.~\ref{dampedHOxis1}. Multiple photons will almost exclusively measure the system's response function, i.e. neither symmetric nor normal ordering but the antisymmetric operator order, $C^{\text{det}}_{\alpha}\gg C^{\text{sym}}$ and the direct symmetric system measurement is negligible. 

\begin{figure}[t]
    \centering
\includegraphics[width=8.5cm]{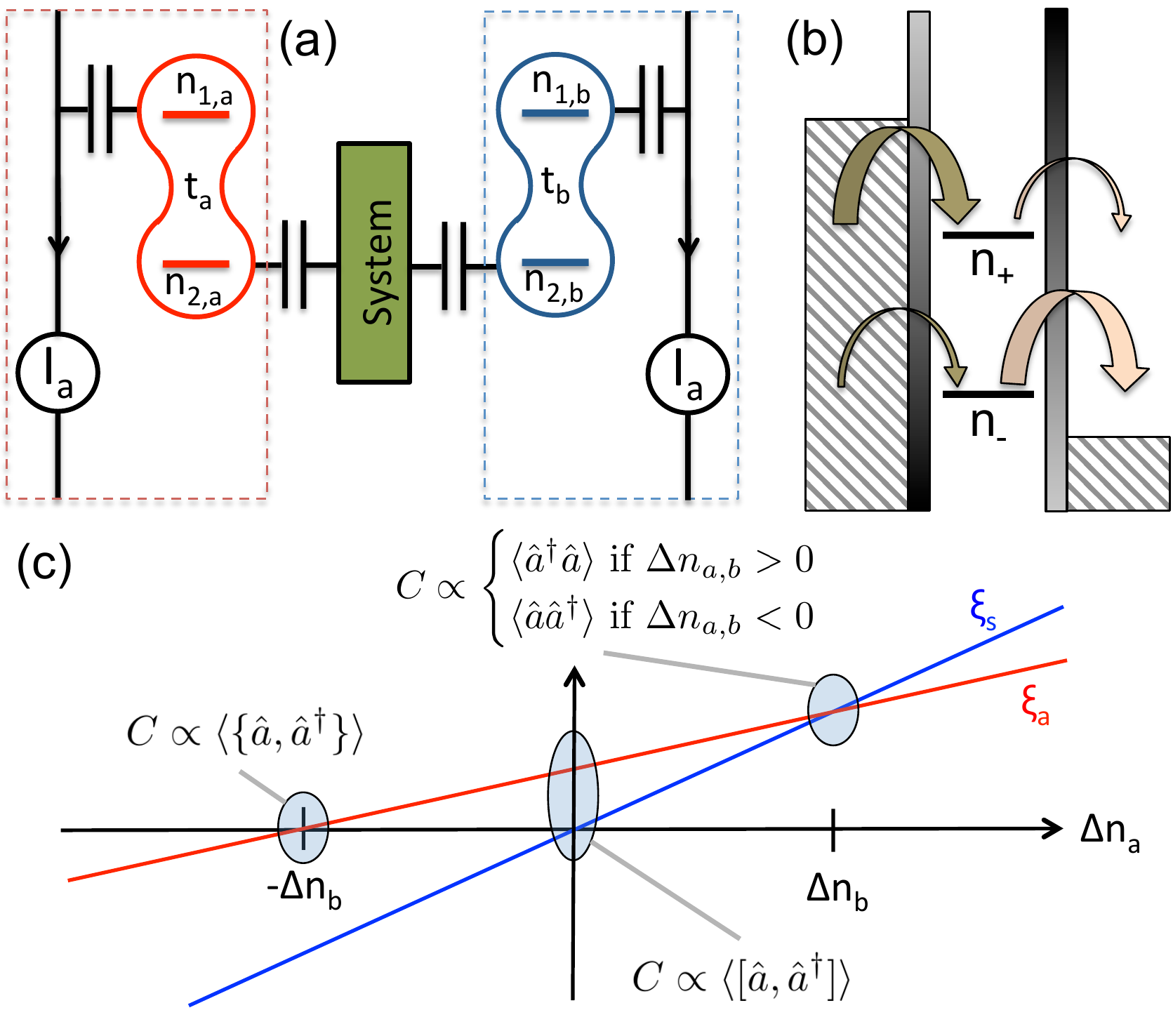}
    \caption{Sketch of the setup. (a) Two double-dot detectors are
      capacitively coupled to a system and the occupation of one of
      the dots is recorded via a bypassing current. (b) The state of
      each detector can be set by additional contacts (not shown in
      (a)) which are biased to adjust the occupation difference of the
      high and low energy eigenstate $n^{+}$ and $n^{-}$. (c) Weights
      $\xi_s$ and $\xi_a$ as a function of the occupation difference
      of the energy eigenstates $\Delta
      n_{\alpha} = n^{-}_{\alpha}-n^{+}_{\alpha}$, $\epsilon_\alpha=0$,
      no damping.} 
     \label{dddsketch}
\end{figure}

\textit{(iii) Detector engineering:}
We now use the possibility to prepare the two
detectors in different states. We consider a
mesoscopic electronic realization with two ($\alpha=a,b$)
double quantum-dot detectors capacitively coupled to the system as depicted in Fig.~\ref{dddsketch}(a). The two dots in each
detector are connected via
a tunnel coupling $t_\alpha$ and the energy difference of the left and
right level $\epsilon_\alpha$ can be tuned via a gate voltage. Both double dots are capacitively connected to a
quantum point contact in which the current $I_{\alpha}$ is
measured. We assume the current correlation function
$\langle I_{a}(t)I_{b}(t')\rangle$ to be proportional to the
occupation number correlation function
$\langle \sigma_{z}^{a}(t)\sigma_{z}^{b}(t')\rangle$ with
$\sigma_z^{\alpha}=n_{1,\alpha}-n_{2,\alpha}$.

The state of the double dots is controlled via a bias voltage, see
Fig.~\ref{dddsketch}(b). It can be characterized by the occupation
difference of the energy eigenlevels
$\Delta n_{\alpha}=n_{\alpha}^{-}-n_{\alpha}^{+}$. By tuning
$\Delta n_\alpha$ from positive to negative values, i.e. setting a
level inversion, the detector can be switched from absorption to
emission mode. By using the cross-correlation of two detectors one can
combine these modes to obtain the symmetrized noise not
accessible to a single detector in either mode.  The rates in and out of
the dot are assumed to be low enough to allow the detector to evolve
undisturbed during the measurement, i.e. we assume the tunneling rates
to determine the initial state and the measurement taking place
between two successive tunneling events.  The detector Hamiltonian is
given by
$\hat
H_{\alpha}=\epsilon_{\alpha}\hat\sigma_z^{\alpha}+t_{\alpha}\hat\sigma_x^{\alpha}$,
and the interaction Hamiltonian reads
$\hat H_{\text{int}}=\sum_\alpha\eta_\alpha \hat\sigma_z^{\alpha}\hat
A$, where both detectors measure the same system variable $\hat A$.
This leads to the response function
\begin{align*}
\chi^{\alpha}_{\sigma_z\sigma_z}(t,t')=&-\theta(t-t')\frac{8 t_\alpha^2}{\omega_\alpha^2}\sin(\omega_\alpha (t-t'))\Delta n_\alpha
\end{align*}
and the noise
\begin{align*}
S^\alpha_{\sigma_z\sigma_z}(t,t')=&\frac{8t_\alpha^2}{\omega_\alpha^2}\cos(\omega_\alpha (t-t'))+\frac{8\epsilon_\alpha^2}{\omega_\alpha^2}\left(1-(\Delta n_\alpha)^2\right)\:,
\label{dddnoise}
\end{align*} 
with $\omega_\alpha=2\sqrt{t_\alpha^2+\epsilon_\alpha^2}$. For
measuring the ladder operators of a harmonic oscillator we can again
write the correlator in the form of Eq.~(\ref{exosc}). The results for
$\xi_s$ and $\xi_a$ are shown in Fig.~\ref{dddsketch}(c). When both
detectors are set to the same occupation difference $\Delta
n_{\alpha}=n_{\alpha}^{-}-n_{\alpha}^{+}$ the noise contributions of
both detectors sum up and we obtain normal order (anti-normal if both
detectors are switched to emission mode) near resonance and for
sufficient low damping. This regime was already explored in Fig.~\ref{dampedHOxis1}. Since
$C^{\text{det}}_\alpha\propto \Delta n_\alpha$, tuning one of the detectors
to level inversion of the other $\Delta n_a=-\Delta n_b$ results in a
cancellation of the terms $C^{\text{det}}_{\alpha}$ and we are able to
measure the pure symmetrized noise/correlation of the system.
If one of the $\Delta n_\alpha$ equals zero, not only the corresponding
$C^{\text{det}}_{\alpha}$ but also $S^{\text{sym}}$ vanishes. In this case the only
remaining contribution to $C$ is the linear-response measurement of
this detector's noise by the other detector via the system, i.e., $C$
only contains information about the antisymmetric
operator order. Tuning the voltage that controls $\Delta n_\alpha$ any
of these regimes can be selected.

A regime to measure the antisymmetric system contribution completely independent of the system state can be explored by utilizing the constant term in the noise. In the frequency domain $C^{\text{det}}_\alpha(\omega=0)$ will be directly proportional to the zero-frequency detector noise $S^{\alpha}_{MD}(\omega=0)=8(\epsilon_\alpha/\omega_\alpha)^2(1-(\Delta n_\alpha)^2)$
which is tunable over a wide range by the gate voltage that controls
$\epsilon_{\alpha}$. If we set a sufficiently large level difference $\epsilon$ the
measurement outcome will be dominated by the detectors measuring each
other's noise through the system. In this way we obtain the pure
antisymmetric system operator order incorporated in $\chi_{AB}^{0}(\omega)$.

\textit{(iv) Markovian limit:} We now consider the
correlation in the limit of short coupling times, $\eta_\alpha\to\delta(t-t_\alpha)$ in Eq.~(\ref{Hint}), in a
stationary setup
\begin{align}
C^{\text{sym}}(t_a,t_b)=&\chi^a_{MD}(0)\chi^b_{MD}(0)S_{AB}^{0}(t_a-t_b)\nonumber\\
C^{\text{det}}_a(t_a,t_b)=&\chi^b_{MD}(0)S_{MD}^{a}(0)\chi^0_{BA}(t_b-t_a)\nonumber\\
C^{\text{det}}_b(t_a,t_b)=&\chi^a_{MD}(0)S_{MD}^{b}(0)\chi^0_{AB}(t_a-t_b)\:.
\end{align}
Note that only one of the terms $C^{\text{det}}_\alpha$ will contribute
depending on whether $t_a>t_b$ or $t_b>t_a$. For thermal detectors,
$\chi^\alpha_{MD}(0)$ and $S_{MD}^{\alpha}(0)$ are related via the
fluctuation-dissipation theorem and one finds a general requirement on
the detectors for measuring the symmetrized correlation.
\begin{figure}[t]
    \centering
\includegraphics[width=7.5cm]{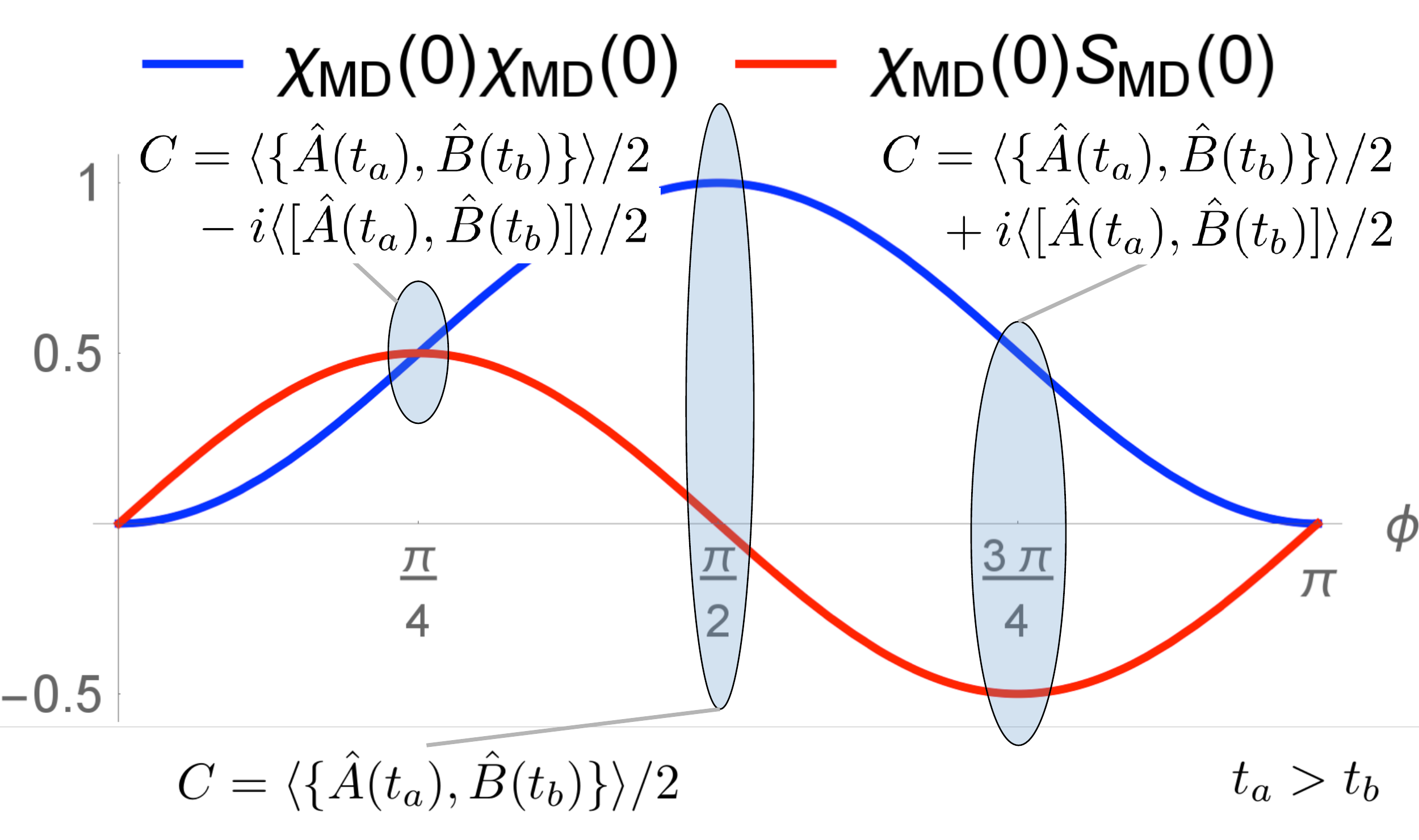}
    \caption{Markovian coupling: The prefactors of $S_{AB}^{0}$ and $\chi^0_{AB}$ as a function of the angle $\phi$ for $\langle\hat\sigma_z\rangle=1$ and $\langle\hat\sigma_x\rangle=\langle\hat\sigma_y\rangle=0$.}
     \label{MarkovianM}
\end{figure}
If $\chi_{MD}^{\alpha}$ (without the Heaviside function) is symmetric in
$t$, e.g. in example \textit{(i)}, $S_{MD}^{\alpha}$ is antisymmetric
in $t$ and therefore $C^{\text{det}}_{\alpha}=0$ and $C$ reduces to the
symmetrized expression. This is independent of the frequency-filter
effect of the detector or any system properties. However, any
antisymmetric contribution in $\chi_{MD}^{\alpha}$ will cause a finite
value of $C^{\text{det}}_\alpha$. For non-thermal detector states there are even less restrictions for $C^{\text{det}}_{\alpha}$ to remain in the Markovian limit.

To be concrete, we consider two identical ($a=b$) two-level
detectors. The detector variable is set to $\hat D=\hat \sigma_x$ and
the detector observable points in direction
$\hat M=\vec{r} \vec{\hat \sigma}$ with
$\vec{r}=(\cos(\phi),\sin(\phi),0)^{T}$. We obtain
$\chi_{MD}(0)=-\sin(\phi)\langle\hat\sigma_z\rangle$ and
$S_{MD}(0)=\cos(\phi)-\langle\vec{r}\vec{\hat
  \sigma}\rangle\langle\hat\sigma_x\rangle$.  Figure~\ref{MarkovianM}
shows that by tuning the angle $\phi$ we can realize the full range of
combinations of the symmetric and antisymmetric operator orders. E.g,
in an optical system as investigated in \textit{(i)} the three points
highlighted in Fig.~\ref{MarkovianM} correspond to measurements of the
$Q$-function, the Wigner function, and the $P$-function.

{\em Conclusion and Outlook.} --- In contrast to theoretical models
which assume instantaneous measurements, constantly coupled detectors
appear naturally in many experiments. In our work we include the
non-Markovian effects in such a setup and explore a variety of
possible outcomes corresponding to non-symmetrized operator orders.
Our results open the avenue to design tunable quantum detection
systems which can observe tailored correlation functions.

We have proposed a quantum detector consisting of a pair of double
quantum dots which realizes such a tunable scheme in a mesoscopic
nanostructured circuit. Applying our analysis to develop quantum
detector setups in the ultrafast optical domain
\cite{rieck:15,moskalenko:15,rieck:17} or in the optical detection of
coherence \cite{trushin:17} appears to be a promising research
direction.  Furthermore, a weak detection scheme is potentially useful
to perform a weak quantum process tomography, which might present an
alternative to the standard route to test quantum algorithms
\cite{Poyatos1997}.  Another interesting future challenge will be to
explore the full statistics of quantum systems in a non-Markovian
detection scheme, eventually even going beyond the weak-measurement
limit.

\begin{acknowledgments}
{\em Acknowledgments.} --- We would like to thank P.P. Hofer for
valuable discussions. This work was financially supported by the
Center of Applied Photonics, an ERC Advanced Grant UltraPhase of
Alfred Leitenstorfer, the DFG through SFB 767, the Swiss SNF, and the
NCCR Quantum Science and Technology.
\end{acknowledgments}

\end{document}